\documentclass[a4paper,fleqn,usenatbib,useAMS]{mnras}

\usepackage{graphicx}	
\usepackage{amsmath}	
\usepackage{amssymb}	
\usepackage{multicol}        
\usepackage{bm}		
\usepackage{pdflscape}	

\newcommand{\tgw}{\ensuremath{\tau_\mathrm{GW}}}
\newcommand{\tem}{\ensuremath{\tau_\mathrm{EM}}}

\usepackage[T1]{fontenc}
\usepackage{ae,aecompl}

\title[SLSN magnetar ellipticity]{
Constraining the ellipticity of strongly magnetized neutron stars powering superluminous supernovae
}

\author[T. J. Moriya and T. M. Tauris]{
Takashi J. Moriya$^{1,2}$\thanks{takashi.moriya@nao.ac.jp}
and
Thomas M. Tauris$^{3,2}$\thanks{tauris@astro.uni-bonn.de}
\\
$^{1}$Division of Theoretical Astronomy, National Astronomical Observatory of Japan, 2-21-1 Osawa, Mitaka, Tokyo 181-8588, Japan \\
$^{2}$Argelander Institute for Astronomy, University of Bonn, Auf dem H\"ugel 71, D-53121 Bonn, Germany \\
$^{3}$Max-Planck Institute for Radio Astronomy, Auf dem H\"ugel 69, D-53121 Bonn, Germany
}

\date{Accepted 2016 April 14. Received 2016 April 14; in original form 2016 March 24}

\pubyear{2016}

\begin{document}
\label{firstpage}
\pagerange{\pageref{firstpage}--\pageref{lastpage}}
\maketitle

\begin{abstract}
Superluminous supernovae (SLSNe) have been suggested to be powered by strongly magnetized, rapidly rotating neutron stars which are 
often called magnetars. In this process, rotational energy of the magnetar is radiated via magnetic dipole radiation and heats the supernova ejecta.
However, if magnetars are highly distorted in their geometric shape, rotational energy is mainly lost as gravitational wave radiation 
and thus such magnetars cannot power SLSNe. By simply comparing electromagnetic and gravitational wave emission timescales, we constrain 
upper limits to the ellipticity of magnetars by assuming that they power the observed SLSNe. 
We find that their ellipticity typically needs to be less than about a~few~$10^{-3}$. This indicates that the toroidal magnetic field strengths 
in these magnetars are typically less than a~few~$10^{16}\;{\rm G}$ so that their distortions remain small. 
Because light-curve modelling of SLSNe shows that their dipole magnetic field strengths are of the order of $10^{14}\;{\rm G}$, 
the ratio of poloidal to toroidal magnetic field strengths is found to be larger than $\sim\!0.01$ in magnetars powering SLSNe.
\end{abstract}

\begin{keywords}
supernovae: general -- gravitational waves -- stars: magnetic fields -- stars: neutron 
\end{keywords}



\section{Introduction}\label{sec:introduction}
Superluminous supernovae (SLSNe) are a newly recognized class of supernovae (SNe) 
whose peak luminosity is more than about 10 times brighter than that of typical core-collapse SNe \citep{quimby2011}. 
These powerful transients are therefore observable at cosmological distances although they are thought to be 
intrinsically rare -- less than $\sim$0.01\% of the core-collapse population \citep{quimby2013,mccrum2015}.
How SLSNe achieve such high luminosities is not well-understood \citep{gal-yam2012}. 
One of the favoured models to explain their peak luminosity is the formation of a rapidly spinning, strongly magnetized neutron star (NS), 
which is often called a `magnetar' in the literature. If such a magnetar has an initial magnetic dipole field strength of $\sim 10^{14}~\mathrm{G}$ 
and an initial spin period of $\sim 1~\mathrm{ms}$, a huge rotational energy reservoir can be emitted as magnetic dipole radiation to heat SN ejecta, 
thus making a SN superluminous \citep{kasen2010,woosley2010,dessart2012,inserra2013,chatzopoulos2013,metzger2015,sukhbold2016}.

On the other hand, magnetars are also suggested to be strong sources of gravitational waves \citep{cutler2002,stella2005,dallosso2009,gualtieri2011,dall'osso2015}. 
The main reason for this is a geometric distortion of the NS caused by a strong toroidal magnetic field. Rapid rotation of a time-varying 
quadrupole moment results in efficient emission of gravitational waves. If SLSNe are actually powered by magnetars, however, 
the released spin-down energy from magnetars must be dominated by electromagnetic radiation which can be thermalized in the SN ejecta.
Therefore, as gravitational waves cannot power the SN ejecta, rotational energy loss by gravitational waves must be insignificant. 
A resulting criterion should be that the spin-down timescale by magnetic dipole radiation must be shorter than that by gravitational wave radiation \citep[cf.][]{kashiyama2015}.

Although magnetars are suggested to be distorted by their strong toroidal magnetic fields, it is not easy to constrain this distortion 
and its resulting ellipticity. Furthermore, the toroidal magnetic field components themselves are also hard to determine observationally \citep{makishima2014,lasky2015}. 
An advantage of using SLSNe to constrain NS ellipticities is that we can use their light curves to estimate the spin period and the 
dipole magnetic field strength of these magnetars. Assuming that magnetars are indeed distorted by toroidal magnetic fields, 
the derived limits on their ellipticities can be used to constrain the toroidal magnetic field strengths in these NSs 
as an independent sanity check. Thus, we can constrain both poloidal and toroidal magnetic field strengths in magnetars in SLSNe 
via modelling of their light curves. 
In this Letter, we show such constraints obtained from SLSNe for which light-curve modelling by magnetars has been performed.

Constraining both poloidal and toroidal field strengths in SLSNe also serves as an important test of the magnetar model of SLSNe. 
It is known that purely poloidal magnetic fields in NSs are not stable, and toroidal magnetic fields which are much stronger than the poloidal fields 
are still required to have stable and significant poloidal magnetic fields \citep[e.g.][]{braithwaite2009,akgun2013}. 
By constraining both poloidal and toroidal field strengths in magnetars powering SLSNe, we can check if these NSs actually satisfy the stability condition, 
and thus, if the magnetar model of SLSNe is self-consistent. 

\section{Constraining the ellipticity}\label{sec:ellipticity}
\subsection{Emission timescales}

We consider a magnetar with a radius $R$, a moment of inertia $I$, and an initial spin period $P_0$. The magnetar has a total rotational energy of 
$E_{\rm rot}=1/2\,I\,(2\pi/P_0)^2$. Furthermore, we idealize the magnetar as a NS with the shape of a slightly deformed, homogeneous ellipsoid 
and thus having a small ellipticity of $\varepsilon \equiv (I_1-I_2)/I_3$, where $I_1$, $I_2$ and $I_3$ are the principal moments
of inertia of the NS and $I_3$ is assumed to be aligned with the spin axis.
Assuming quadrupolar gravitational wave radiation, the gravitational wave luminosity $L_\mathrm{GW}$ from the distorted magnetar is \citep{st83}:
\begin{equation}
  L_\mathrm{GW}(t) = \frac{E_{\rm rot}}{\tgw}\left(1+\frac{2\,t}{\tgw}\right)^{-3/2}\;,
\label{eq:Lgw}
\end{equation}
where $\tgw \equiv |E_{\rm rot}/\dot{E}_{\rm rot}|$ is the gravitational wave emission timescale of the magnetar:  
\begin{equation}
  \tgw = \frac{5}{2^{10}\pi^4}\frac{c^5P_0^4}{GI\varepsilon^2}\;,
\label{eq:tgw}
\end{equation}
$c$ is the speed of light, and $G$ is the gravitational constant. Here, we assume that the angle between 
the spin axis and the principal axis of the NS distortion is $\pi/2$ \citep{cutler2001}.

In principle, rapidly spinning NSs may emit gravitational waves from a combination of `mountain' and `wobble' radiation,
see e.g. \citet{lasky2015} for a recent discussion on CFS instabilities in the form of f modes (bar-modes) and r modes
(inertial modes, \citealt{ak01}). However, their mode oscillation amplitudes are likely saturated at modest values, resulting in relatively
long spin-down timescales of several years \citep{lasky2015}.

Spinning magnetars with a misaligned magnetic dipole moment emit magnetic dipole radiation with a luminosity $L_\mathrm{EM}$
which is approximately given by \citep{st83}:  
\begin{equation}
  L_\mathrm{EM}(t)=\frac{E_{\rm rot}}{\tem}\left(1+\frac{t}{\tem}\right)^{-2}\;,
\end{equation}
where $\tem \equiv |E_{\rm rot}/\dot{E}_{\rm rot}|=P/2\dot{P}$ is the electromagnetic radiation timescale of the magnetar: 
\begin{equation}
  \tem = \frac{3}{4\pi^2}\frac{Ic^3P_0^2}{B_{\rm dipole}^2R^6\sin^2\alpha}\;,
\label{eq:tem}
\end{equation}
$B_{\rm dipole}$ is the dipole magnetic field strength at the pole, and $\alpha$ is the misalignment angle between the spin axis and the magnetic dipole axis.
We assume $\alpha=\pi/2$.

\subsection{Constraining ellipticity from SLSN light curves}
Long rise times and large luminosities of SLSNe require extraordinary central engines. If magnetars power the observed light curves
they must deposit their rotational energy of more than $10^{51}~\mathrm{erg}$ through electromagnetic radiation with a timescale of 
more than 10~days \citep{kasen2010}. This requires $B_{\rm dipole}\sim\!10^{14}~\mathrm{G}$ and $P_0\sim\!1~\mathrm{ms}$ (Table~\ref{table:NSepsilon}). 
However, on the other hand, if the gravitational wave emission timescale is shorter than the electromagnetic emission timescale, 
the rotational energy of the magnetar is mainly lost by gravitational wave radiation and SLSNe cannot be powered by magnetars. 
Thus, we simply impose the condition that the gravitational wave emission timescale needs to be larger than the electromagnetic wave emission timescale 
in SLSNe, i.e., $\tgw > \tem$. Using Equations~(\ref{eq:tgw}) and (\ref{eq:tem}), we obtain the following constraint on the NS ellipticity,
\begin{equation}
  |\varepsilon| < \,\sqrt[]{\frac{5}{3G}}\frac{cR^3P_0B_{\rm dipole}}{2^4\pi I}\simeq3.0\times 10^{-4}\;\left(\frac{B_{\rm dipole}}{10^{14}~\mathrm{G}}\right)
                  \left(\frac{P_0}{1~\mathrm{ms}}\right),
\label{eq:elipticityconstraint}
\end{equation}
where we apply fiducial NS properties of $I=10^{45}~\mathrm{g~cm^2}$ and $R=10~\mathrm{km}$.

\begin{figure}
 \begin{center}
  \includegraphics[width=0.74\columnwidth,angle=-90]{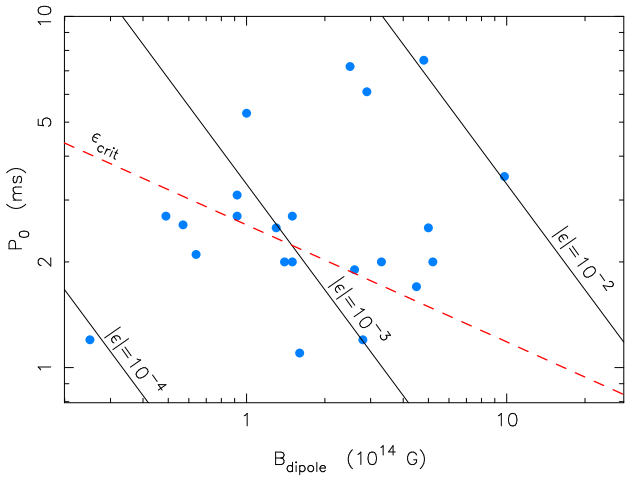}
 \end{center}
\caption{
Initial dipole field strengths and spin periods of magnetars in observed SLSNe (blue points, see Table~\ref{table:NSepsilon}). 
The corresponding constraints on ellipticity (Equation~\ref{eq:elipticityconstraint}) are shown with three solid black lines. 
Spin flipping in magnetars can occur in magnetars below the dashed red line \citep{lasky2015}.
}\label{fig:bpdiagram}
\end{figure}

Table~\ref{table:NSepsilon} summarizes the estimated magnetar properties ($B_{\rm dipole}$ and $P_0$) for observed SLSNe 
and their corresponding constraints on the NS ellipticity obtained with Equation~(\ref{eq:elipticityconstraint}). 
The angle $\alpha$ is defined differently depending on the reference in Table~\ref{table:NSepsilon}. We take $B_{\rm dipole}\sin\alpha$ in the literature and obtain $B_{\rm dipole}$ by assuming $\alpha=\pi/2$ (see Equation~\ref{eq:tem}).
Figure~\ref{fig:bpdiagram} depicts these values graphically. 
We can see that the absolute NS ellipticity typically needs to be less than 
$\sim\!10^{-3}$ for electromagnetic wave radiation to be more efficient than gravitational wave radiation in the observed SLSNe.

\begin{table*}
\begin{tabular}{lccccl}
\hline
SN name & $B_{\rm dipole}$ & $P_0$ & $|\varepsilon|$ &$B_{\rm toroidal}$& Reference \\
 & $10^{14}\mathrm{G}$ & ms & $10^{-3}$&$10^{16}~\mathrm{G}$ & \\
\hline
SN 2005ap& 0.92 & 3.1 & $<0.85$& $<2.4$ &\citet{chatzopoulos2013}\\
SCP06F6  & 1.3 & 2.5 & $<0.98$ & $<2.5$ &\citet{chatzopoulos2013}\\
SNLS 06D4eu&1.4& 2.0 & $<0.85$ & $<2.3$ &\citet{howell2013}\\
SN 2007bi& 0.92 & 2.7 & $<0.75$& $<2.2$ &\citet{chatzopoulos2013}\\
SN 2010gx& 5.2 & 2.0 & $<3.1$  & $<4.5$ &\citet{inserra2013}\\
SN 2010kd& 1.5 & 2.7 & $<1.2$  & $<2.8$ &\citet{chatzopoulos2013}\\
SN 2010kl& 9.8& 3.5 & $<10 $  & $<8.2$ &\citet{bersten2016}\\
PTF10hgi & 2.5 & 7.2 & $<5.4$  & $<5.9$ &\citet{inserra2013}\\
SN 2011ke& 4.5 & 1.7 & $<2.3$  & $<3.9$ &\citet{inserra2013}\\
SN 2011kf& 3.3 & 2.0 & $<2.0$  & $<3.6$ &\citet{inserra2013}\\
PTF11rks & 4.8 & 7.5 & $<11 $  & $<8.4$ &\citet{inserra2013}\\
SN 2012il& 2.9 & 6.1 & $<5.3$  & $<5.9$ &\citet{inserra2013}\\
PTF12dam & 0.49& 2.7 & $<0.39$ & $<1.6$ &\citet{chen2015}\\
CSS121015& 1.5 & 2.0 & $<0.90$ & $<2.4$ &\citet{nicholl2014} \\
LSQ12dlf & 2.6 & 1.9 & $<1.5$  & $<3.1$ &\citet{nicholl2014} \\
SSS120810& 2.8 & 1.2 & $<1.0$  & $<2.6$ &\citet{nicholl2014} \\
SN 2013dg& 5.0 & 2.5 & $<3.7$  & $<5.0$ &\citet{nicholl2014} \\
iPTF13ajg& 1.6 & 1.1 & $<0.54$ & $<1.9$ &\citet{vreeswijk2014} \\
iPTF13ehe& 0.57& 2.55& $<0.43$ & $<1.7$ &\citet{wang2015} \\
DES13S2cmm& 1.0& 5.3 & $<1.6$  & $<3.2$ &\citet{papadopoulos2015} \\
SN 2015bn & 0.64& 2.1 & $<0.40$& $<1.6$ &\citet{nicholl2016} \\
ASASSN-15lh&0.25 &1.2 & $<0.090$ & $<0.77$&\citet{bersten2016} \\
\hline
\end{tabular}
\caption{Constraints on the ellipticity and toroidal magnetic field component in magnetars applied to observed SLSNe. We assume $\alpha = \pi/2$.}
\label{table:NSepsilon}
\end{table*}

\section{Discussion}\label{sec:discussion}
If we assume that a NS distortion is indeed caused by a strong internal toroidal magnetic field component, $B_{\rm toroidal}$ we can constrain 
the average value of this component by using a relation between $\varepsilon$ and $B_{\rm toroidal}$. If we take the suggested relation
by \citet{cutler2002}, $|\varepsilon|\simeq 1.6\times 10^{-4}(B_{\rm toroidal}/10^{16}~\mathrm{G})^2$, and combine with Equation~(\ref{eq:elipticityconstraint}),
we can constrain $B_{\rm toroidal}$ as: 
\begin{equation}
  B_{\rm toroidal} \lesssim  1.4\times 10^{16}~\mathrm{G}\,\left(\frac{B_{\rm dipole}}{10^{14}~\mathrm{G}}\right)^{1/2}\left(\frac{P_0}{1~\mathrm{ms}}\right)^{1/2}\;.
\label{eq:toroidalconstraint}
\end{equation}
The resulting constraints on $B_{\rm toroidal}$ are shown in Table~\ref{table:NSepsilon} and are typically less than a~few~$10^{16}$~G. 
This is close to the maximum toroidal magnetic field strength which can be achieved by the $\alpha-\Omega$ dynamo mechanism \citep{duncan1992,moesta2015}.

To secure a stable magnetic configuration, one must require $B_{\rm dipole}\ll B_{\rm toroidal}$ \citep{braithwaite2009,akgun2013}. 
Here, we have constraints on both magnetic field components from fitting a magnetar to the light curves of the SLSNe (Table~\ref{table:NSepsilon}) and
the ratio of the two components is typically found to be $B_{{\rm dipole}}/B_{\rm toroidal}\gtrsim\!0.01$. 
Hence, the criterion for a stable magnetic field configuration is validated and our derived magnetar properties, obtained from the 
applied magnetar model for SLSNe, are self-consistent (but see also \citealt{soker2016}). If future observations of SLSNe yield a much larger ratio of $B_{{\rm dipole}}/B_{\rm toroidal}$ 
when applying the magnetar model it may well indicate that magnetars, at least in those cases, are not related to SLSNe.

Our estimate of the gravitational wave radiation timescale (Equation~\ref{eq:tgw}) is based on the assumption that the angle between the spin axis 
and the principal axis of the NS distortion is $\pi/2$. Furthermore, if this distortion in the NS quadrupole moment is caused by toroidal magnetic fields, 
then these fields should be orthogonalized to the spin axis \citep{cutler2001}. However, when magnetars are formed and their toroidal fields are wound up, their toroidal fields are presumed to be aligned, 
and not orthogonal, to the spin axis \citep{duncan1992}. Therefore, if orthogonalization does not occur, then the toroidal magnetic fields are not able to 
distort the NSs sufficiently to become efficient gravitational wave emitters. The timescale for this orthogonalization, or `spin flipping', 
is quite uncertain \citep{dallosso2009}. \citet{lasky2015} showed that a critical ellipticity exists for newborn magnetars 
such that they will only cause spin flipping if $|\varepsilon| < \varepsilon_{\rm crit} \simeq 5\times 10^{-3}(P_0/1~\mathrm{ms})^{-2}$ for typical values 
of the NS mass density ($10^{15}~\mathrm{g~cm^{-3}}$) and radius (10~km). 
Applying Equation~(\ref{eq:elipticityconstraint}), we obtain the dashed red line in Figure~\ref{fig:bpdiagram}, below which the estimated ellipticities  
are smaller than the critical ellipticity. Hence, in all observed sources above this red dashed line, the magnetars might always be
inefficient gravitational wave emitters and all their rotational energy loss can, in principle, be used to power the observed SLSN light curves. 
Sources below the red dashed line, can still power SLSNe if their ellipticities are smaller than the derived upper limits shown in Table~\ref{table:NSepsilon}. 
There are large uncertainties, however, in estimating the spin-flip criterion, and further investigation is needed on this issue.

A slight caveat of concern for our estimated description of $\tem$ is related to the validity of applying the simple magnetic dipole model
to a magnetar in a SN. Such a magnetar is surrounded by an expanding envelope and its braking torque depends on  
the boundary conditions at the wind-envelope interface. If initially the B-field does not penetrate the envelope, 
then the spin-down torque will be smaller \citep{lb03}; if it penetrates, it will be larger (since the field is twisted more and more). We also note that numerical magnetohydrodynamic simulations indicate that the spin-down luminosity may be larger than that obtained by the classical dipole formula by a factor of $\sim 2$ \citep[e.g.][]{spitkovsky2006,tchekhovskoy2013}.

In addition, there are uncertainties in estimating $B_\mathrm{dipole}$ and $P_0$ from SLSN light curves. These magnetar properties are often entangled with SN properties such as ejecta mass, energy and opacity. This degeneracy can be partly solved by using the velocity information from spectra, but still yields $B_\mathrm{dipole}\sim 10^{14}~\mathrm{G}$ and $P_0\sim 1~\mathrm{ms}$ \citep[e.g.][]{nicholl2015}. Thus, we expect that $B_\mathrm{dipole}$ and $P_0$ in Table~\ref{table:NSepsilon} have uncertainties by a factor of a few. We also note that the efficiency in converting magnetar dipole radiation to thermal energy can also affect the estimates of the magnetar properties \citep{chen2015}.

The ellipticities and spins of NSs at birth are not well constrained, especially not for magnetars. 
A recent study on the evolution of proto-NSs \citep{camelio2016} suggests a minimum proto-NS spin period of about 3~ms 
(obtained some 10~s after core bounce), and thus significantly larger than the mass-shedding limit of $\sim\!0.6-0.7\;{\rm ms}$ \citep{dgkk13}. 
However, several of the SLSNe investigated here require $P_0<3\;{\rm ms}$. Therefore, if proto-NSs (magnetars) 
are actually hard to spin up efficiently, then these SLSNe cannot be powered by magnetars.

There are several observational constraints on ellipticities of more evolved magnetars. 
\citet{makishima2014} estimated that 
the ellipticity of the magnetar 4U~0142+61 with $B_\mathrm{dipole}\sim\!10^{15}~\mathrm{G}$ is $1.6\times 10^{-4}$. 
If this deformation is due to a toroidal magnetic field component, the corresponding toroidal magnetic field strength is $\sim\!10^{16}~\mathrm{G}$. 

For NS mergers producing short gamma-ray bursts, the resulting (meta-stable) magnetar is significantly more massive 
($>2.2\;M_{\odot}$) than a typical NS ($1.4\;M_{\odot}$). Hence, these sources can be efficient gravitational wave emitters via f-modes \citep{dkp15}.
\citet{lasky2015} estimated that magnetars with $B_\mathrm{dipole}\gtrsim 10^{15}~\mathrm{G}$ powering short gamma-ray bursts, 
have ellipticities of less than $\sim\!10^{-2}$. In comparison, we find that magnetars potentially powering SLSNe have about an order of magnitude 
smaller values of $B_{\rm dipole}$ and the magnetar ellipticities that we derived are typically less than a~few~$\sim\!10^{-3}$.
Therefore, if magnetars power SLSNe then we do not expect to detect these (extragalactic) NSs by any of the current (aLIGO/VIRGO/KAGRA) nor planned gravitational wave 
observatories (Einstein Telescope), as their gravitational wave signals must be quite weak.
We refer to \citet{kashiyama2015} for further discussion on the gravitational wave detectability from SLSNe.

\section{Conclusions}
We have derived constraints on the ellipticity of magnetars powering SLSNe. For magnetars to power SLSNe, 
electromagnetic radiation should be the dominant channel to extract their large rotational energy reservoir. Thus, the magnetar 
ellipticity must be small in order to prevent significant loss of rotational energy by gravitational wave radiation from 
geometrically distorted magnetars. Here, we simply constrained the ellipticity by requiring that the electromagnetic radiation timescale 
should be shorter than gravitational wave radiation timescale in distorted magnetars.

We find that the magnetar ellipticity, $\varepsilon$ in SLSNe typically needs to satisfy $|\varepsilon| \lesssim \mathrm{a~few}\times 10^{-3}$. 
Thus, their toroidal magnetic field strengths should be smaller than a~few~$10^{16}$~G. 
Combined with the poloidal (dipole) magnetic field strengths constrained by light-curve modelling of SLSNe ($\sim 10^{14}$~G), 
we find that the ratio of poloidal to toroidal field strengths is larger than $\sim\!0.01$ in magnetars powering SLSNe. 
This ratio is small enough to secure stable magnetic configuration in magnetars powering SLSNe and thus the magnetar model for SLSNe 
is found to be self-consistent so far. 

\section*{Acknowledgements}
This work is based on discussions following a presentation by TJM at the \textit{8th BONN workshop on formation and evolution of neutron stars}.
We thank Nils Andersson and Kostas Kokkotas for discussions on r-mode and f-mode instabilities of magnetars, and Maxim Lyutikov for insight
on the spin-down of a magnetar in a SN.
TJM is supported by Japan Society for the Promotion of Science Postdoctoral Fellowships for Research Abroad (26\textperiodcentered 51).










\bsp	
\label{lastpage}
\end{document}